\documentclass[prb,twocolumn,superscriptaddress,showpacs,floatfix]{revtex4-1}  
\usepackage{amsmath}
\usepackage{color,soul}
\usepackage{txfonts}
\usepackage{mathrsfs}
\usepackage{dcolumn}
\usepackage{bm}
\usepackage{graphicx}
\usepackage{multirow}

\usepackage{xfrac}
\allowdisplaybreaks

\begin{document}

\title{Photoenhanced spin/valley polarization and tunneling magnetoresistance in ferromagnetic-normal-ferromagnetic silicene junction}
\date{\today}
\author{Le Bin Ho}
\email{binho@qi.mp.es.osaka-u.ac.jp}
\affiliation{Ho Chi Minh City Institute of Physics, VAST, Ho Chi Minh City, Vietnam}
\affiliation{Graduate School of Engineering Science, Osaka University, Toyonaka, Osaka 560-8531, Japan}  
\author{Tran Nguyen Lan}
\email{latran@umich.edu}
\affiliation{Ho Chi Minh City Institute of Physics, VAST, Ho Chi Minh City, Vietnam}
\affiliation{Department of Physics, University of Michigan, Ann Arbor, Michigan, 48109, USA}

\begin{abstract}
We theoretically demonstrate a simple way to significantly enhance the spin/valley polarizations and tunneling magnetoresistnace (TMR) in a ferromagnetic-normal-ferromagnetic (FNF) silicene junction by applying a circularly polarized light in off-resonant regime to the second ferromagnetic (FM) region.
We show that the fully spin-polarized current can be realized in certain ranges of light intensity.
Increasing the incident energy in the presence of light will induce a transition of perfect spin polarization from positive to negative or {\it vice versa} depending on magnetic configuration (parallel or anti-parallel) of FNF junction.
Additionally, under a circularly polarized light, valley polarization is very sensitive to electric field and the perfect valley polarization can be achieved even when staggered electric field is much smaller than exchange field.
The most important result we would like to emphasize in this paper is that the perfect spin polarization and 100\% TMR induced by a circularly polarized light are completely independent of barrier height in normal region.
Furthermore, the sign reversal of TMR can be observed when the polarized direction of light is changed.
A condition for observing the 100\% TMR is also reported.
Our results are expected to be informative for real applications of FNF silicene junction, especially in spintronics.
\end{abstract}
\maketitle
\section{Introduction}

Silicene, a two-dimensional allotrope of Si, has attracted a great attention in both theory and experiment.
This material not only shares all intriguing electronic properties of graphene, but it has some superior advantages compared to graphene, such as strong spin-orbit coupling and buckled honeycomb structure.
While the former enables us to realize the quantum spin Hall effect \cite{Liu2011}, the latter allows us to control the bulk band gap of silicene by applying an external electric field \cite{Drummond2012}.

Recently, Ezawa reported multiple topological phase transitions of silicene in the presence of electric and exchange fields \cite{Ezawa_prl2012,Ezawa_njp2012,Ezawa_prb2013,Ezawa_epjb2012,Ezawa_epjb2012}.
Influences of these fields on ballistic transport of single and multiple barrier junctions of silicene have been also extensively investigated  \cite{Yokoyama_prb2013,Wang_apl2014,Soodchomshom_jap204,Missault_prb2016,Missault_prb2015,Wang_apl2014,Vargiamidis_apl2014,Vargiamidis_jap2015}.
These studies have found many interesting and novel transport phenomena, which is not analogous in graphene, for examples, field-dependent spin- and valley-polarized currents or field-dependent transport gap.

Although a circularly polarized light in off-resonant regime can also produce a topological phase transition in silicene \cite{Ezawa_prl2013}, its effect on ballistic transport has not received a deserving attention and only done by a few recent works.
Tahir and Schwingenschl\"{o}gl \cite{Tahir_epjb2015} studied the Hall and longitudinal conductivities of silicene and germanene in the presence of a perpendicular electric and magnetic fields taking into account the effects of off-resonant light.
Meantime, Niu and Dong \cite{Niu_epl2015} reported effects of off-resonant light in combination with staggered electric field or gate voltage on spin and valley polarizations of a normal silicene junction.
These authors also considered the variation of TMR with the gate voltage applied to third region of a FNF silicene junction when FM regions were exposed to a fixed circularly polarized light.
Nevertheless, a detailed investigation on the influence of circularly polarized light itself on spin and valley polarizations as well as TMR of a FNF silicene junction has not been fully established yet.
Thus, some important effects still remain unexplored.
On one hand, the exchange field in FM regions strongly break the spin degeneracy of band structure.
On the other hand, the circularly polarized light in the off-resonant regime will differently open band gaps of different spin channels, which is unlike in graphene because of strong spin-orbit coupling in silicene.
Consequently, it is expected to obtain intriguing spin-polarized transport phenomena under the interplay between exchange field and circularly polarized light in a FNF silicene junction.

Recent studies have shown that spin and valley polarizations of a FNF silicene junction strongly depend on the barrier height in normal region  \cite{Qiu_jpd2015,Hajati_sumi2016264}.
Furthermore, TMR has an oscillatory behavior with respect to incident energy and barrier height \cite{Wang_jap2013}.
It is therefore non-trivial to obtain a robust polarized current and large TMR value in practice.
A technique to enhance the spin and valley polarizations as well as TMR of a FNF silicene junction is then highly desirable.
Very recently, Saxena {\it et al.} \cite{Saxena_prb2015} theoretically showed that by applying an appropriate external electric field to FM regions, the 100\% positive TMR regardless of barrier $U$ can be found.

In this paper, we systematically investigate the ballistic transport in a FNF silicene junction in the presence of a circularly polarized light in off-resonant regime applied to the third region.
We show that in the presence of light, the sign of perfect spin polarization can be switched by increasing incident energy.
The fully valley-polarized current is very sensitive to the staggered electric potential even when it is much smaller than exchange field.
The most importantly, the perfect spin polarization and 100\% TMR irrespective of barrier height in NM region can be obtained.
Moreover, the sign of TMR can be also fully controlled by the polarized direction of light.
The condition to realize 100\% TMR is also provided.

\section{Theory}

\subsection{Model Hamiltonian}

Let us now shortly introduce the circularly polarized light, $\mathbf{A}(t) = A(\sin(\Omega t), \cos(\Omega t))$, which is perpendicularly applied to the silicene junction.
$\Omega$ is the frequency of light, $\Omega > 0$ for right polarization and $\Omega < 0$ for left polarization.
In this work, we only focus on the {\it off-resonant} regime. Such regime occurs when $\hbar\Omega  \gg t_0$, where $t_0 = 1.6$ eV is the nearest-neighbor hopping in silicene.
In the limit of $eAv_F/\hbar\Omega \ll 1.0$ (with $v_F$ as the Fermi velocity), the off-resonant light is presented {\it via} a static effective Hamiltonian
\begin{align}
  H_{eff} = \eta\lambda_\Omega\tau_z,
\end{align}
where $\lambda_\Omega = (eAv_F)^2/\hbar\Omega$, $\eta$ stands for valley index, and $\tau_i$ (with $i = x,y,z$) are the Pauli matrices of the sublattice pseudospin.
A detailed derivation of $H_{eff}$ can be found in Refs.~\cite{Kitagawa_prb2011,Ezawa_prl2013}.
\begin{figure} [h]
  \includegraphics[width=5.0cm,height=3.0cm]{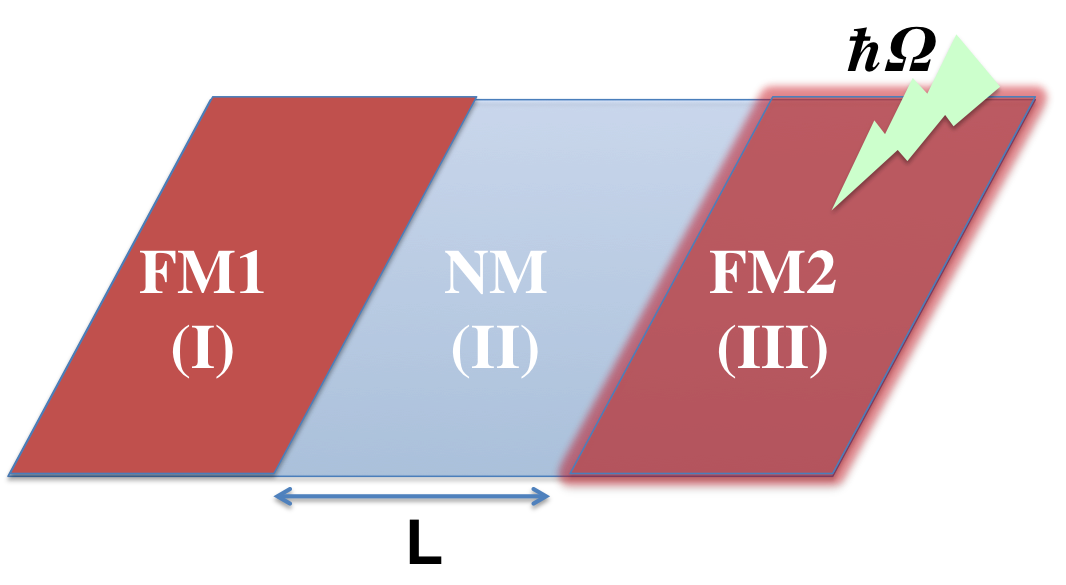}
  \caption{ Schematic picture of a FNF silicene junction, in which the third region (FM2) is exposed to a circularly polarized light.
    $L$ is the length of normal region (NM).}
  \label{fnf_scheme}
\end{figure}

The low-energy effective Hamiltonian of a FNF silicene junction depicted in Fig.~\ref{fnf_scheme} is given by
\begin{align}{\label{Heff}}
  H = \hbar v_F\left(\eta k_x\tau_x+k_y\tau_y\right) + \Delta_{\eta\sigma}\tau_z + U - \sigma h_r,
\end{align}
with $v_F \approx 5.5\times 10^5$ m/s for silicene, $\Delta_{\eta\sigma} = \eta \sigma \lambda_{SO}-\Delta_z+\eta \lambda_\Omega$, $k^2=k_x^2+k_y^2$, and indices $r = 1,\, 2,\, 3$ stand for regions I (FM1), II (NM), and III (FM2), respectively. 
$\lambda_{SO} = 3.9$ meV is the spin-orbit coupling constant in silicene and $\Delta_z$ is the staggered electric potential between $A$ and $B$ sublattices, which is induced by an external electric field perpendicular to the plane of silicene.
$\eta = \pm 1$ corresponds to the $K$ and $K'$ points, and $\sigma = \pm 1$ respectively denotes spin up and down.
$U$ is the barrier potential induced by a gate voltage applied to the normal silicene region.
$h_r$ is the exchange field in FM regions ($h_2 = 0$), where $h_1 = h_3 = h$ and $h_1 = -h_3 = h$ correspond to P and AP configurations.
Note that because the Hamiltonian presented circularly polarized light in off-resonant regime is valley degenerate, the circularly polarized light itself cannot generate a valley-polarized current.
However, thanks to the valley degeneracy, it is expected to provide significant effects on spin-polarized transport properties.
Furthermore, it is easy to recognize that the valley polarization induced by staggered electric field will be significantly enhanced if the junction is additionally exposed to a circularly polarized light.
In this work, the electric field and circularly polarized light are only applied to the third region.

\subsection{Transport calculation}

The general eigenfunctions of Hamiltonian~(\ref{Heff}) are of the form $\psi(x,y) =  (u_A,u_B)^Te^{ik_xx}e^{ik_yy}$.
Writing down 2-component Dirac equation using Hamiltonian (\ref{Heff}) and wavefunction $\psi(x,y)$ for each region, we obtain 

\begin{align}{\label{Dirac}}
\begin{pmatrix}
 E_r-m_r&\hbar v_F(\eta k_x - ik_y) \\
 \hbar v_F(\eta k_x + ik_y)&E_r+m_r		
 \end{pmatrix}
 \begin{pmatrix}
 u_A \\
 u_B	
 \end{pmatrix}
 = 0,
\end{align}
where, $E_1 = E+\sigma h_1,\, E_2 = E-U,\, E_3 = E + \sigma h_3,\, m_1 = m_2 = \eta \sigma \lambda_{SO},\, m_3=\eta \sigma\lambda_{SO} - \Delta_z + \eta\lambda_\Omega$.
Here, $E$ is an incident energy.

We then define transfer matrices $M_r$ as
\begin{align}{\label{matrixM}}
M_r=
\begin{pmatrix}
 e^{iK_X^r\xi}&e^{-iK_X^r\xi} \\
 \dfrac{\eta K_X^r-iK_Y}{E_r+m_r}e^{iK_X^r\xi}&\dfrac{-\eta K_X^r+iK_Y}{E_r+m_r}e^{-iK_X^r\xi}
 \end{pmatrix},
\end{align}
with dimensionless quantity $\xi = x/L$ and
\begin{align}
  K_Y   &=Lk_y, \nonumber \\ 
  K_X^r &= {\rm sign}(E_r+m_r)\sqrt{\left(E_r^2-m_r^2\right)-K_Y^2}. \nonumber
\end{align}
Note that the translational invariance along the $y$ axis has been used.
These transfer matrices $M_r$ will lead to the relation
\begin{align}{\label{uAB}}
  \begin{pmatrix}
    u_A \\
    u_B	
  \end{pmatrix}
  =M_r
  \begin{pmatrix}
    a_r \\
    b_r	
  \end{pmatrix},
\end{align}
where $(a_r,\,b_r)$ are unknown wavefunction coefficients of region $r$.

Let $r_{\sigma\eta}$ and $t_{\sigma\eta}$ be reflection coefficient in region I and transmission coefficient in region III, respectively, we have $(a_1, b_1) = (1, r_{\sigma\eta})$ and $(a_3, b_3) = (t_{\sigma\eta}, 0)$.
We finally obtain the following equation
\begin{align}{\label{matrixT}}
\begin{pmatrix}
1 \\
r_{\sigma\eta}	
 \end{pmatrix}
 =M_1^{-1}M_2M_2^{-1}M_3
 \begin{pmatrix}
t_{\sigma\eta} \\
0	
 \end{pmatrix}.
\end{align}
The transmission probability is then evaluated according to the formula $T_{\sigma\eta}=|t_{\sigma\eta}|^2$.
The normalized spin-valley dependent conductance at zero temperature is evaluated according to Landauer-B\"{u}ttiker formalism as
\begin{align}
  G_{\sigma\eta} = \frac{1}{2}\int_{-\pi/2}^{\pi/2} T_{\sigma\eta} (E,\phi)\cos(\phi)d\phi.
\end{align}
The spin(valley) polarization $P_{s(v)}$  and $TMR$ (in \%) are then defined as follows
\begin{align}{\label{G}}
  P_{s(v)} &=\dfrac{G_{\uparrow(K)}-G_{\downarrow(K')}}{G}, \\
  TMR &= \dfrac{G_P-G_{AP}}{G_P+G_{AP}} \times 100\%,
\end{align}
where the spin(valley)-resolved and total conductances are in turn given by $G_{\sigma(\eta)} = \left(G_{\sigma K ({\uparrow \eta})}+G_{\sigma K'(\downarrow \eta)}\right)/2$ and $G = G_{\uparrow(K)}+G_{\downarrow(K')}$.
$G_P$ and $G_{AP}$ are total conductances of P and AP configurations, respectively.

\section{Numerical results}

Ezawa showed that the lowest frequency to satisfy the off-resonant condition for silicene is 1000 THz \cite{Ezawa_prl2013}.
In this work, we have assumed that the junction is exposed to a circularly polarized light in the soft x-ray regime with high frequency of 4000 THz.
With the maximum value of $\lambda_\Omega$ up to 6.0$\lambda_{SO}$ used herein, the limit $eAv_F/ \hbar\Omega \ll 1.0$ still holds if a strong intensity $I = (eA\Omega)^2/(8\pi\hbar\alpha) \approx 0.4\times 10^{12}$ Wcm$^{-2}$ (with $\alpha = 1/137$) is used, i.e. $eAv_F/ \hbar\Omega \approx 0.09$.

\subsection{Spin and valley polarizations}

\begin{figure} [h]
  \includegraphics[width=7.0cm,height=4.25cm]{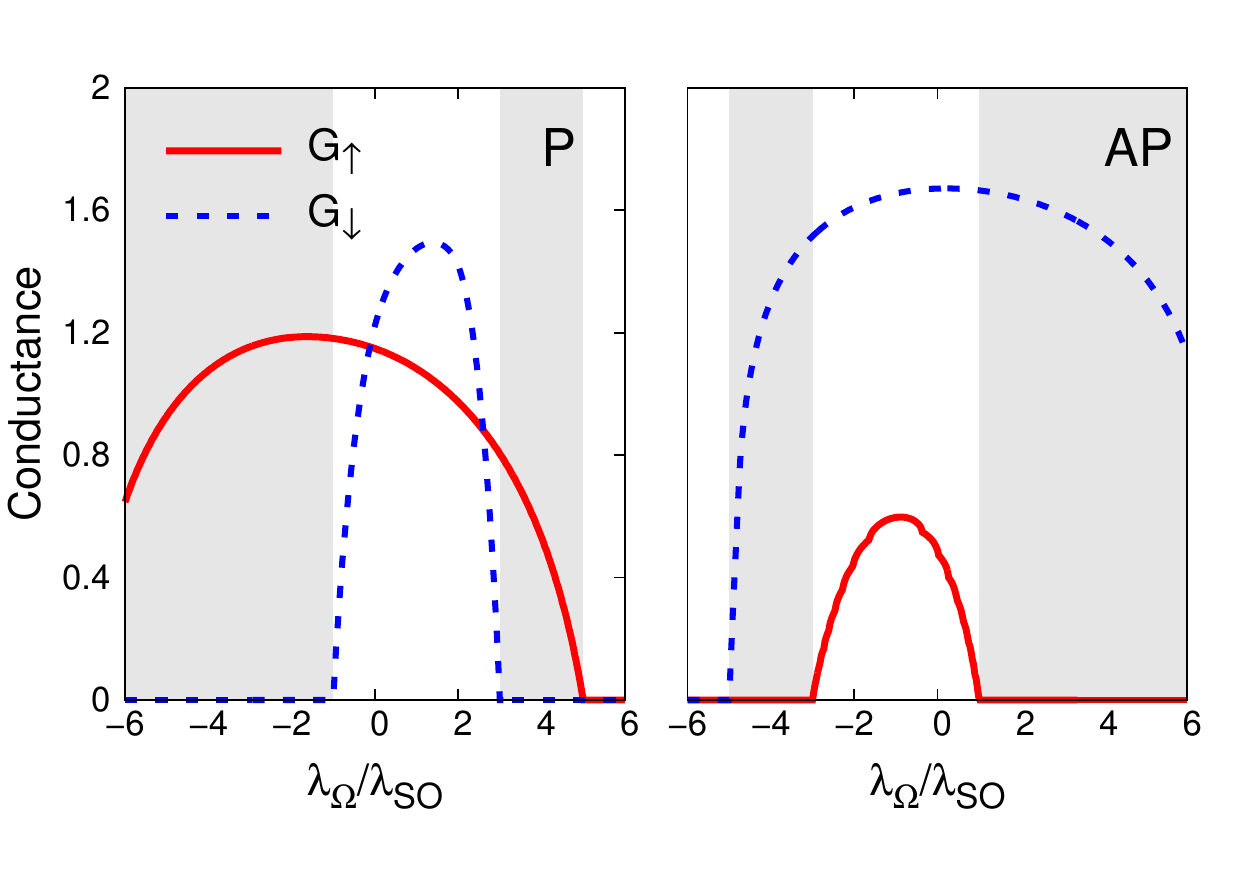}
  \caption{Spin-resolved conductances as functions of $\lambda_\Omega$ for P and AP configurations.
    Parameters used are: $E/\lambda_{SO} = 2.0, h/\lambda_{SO} = 4.0,$ and $U/\lambda_{SO} = 20.0$.}
  \label{fig:GA}
\end{figure}

First, we will analyze the spin polarizations of both P and AP configurations under the presence of circularly polarized light.
The staggered electric potential will not be included ($\Delta_z = 0$) to fully explore the role of circularly polarized light.
Fig. \ref{fig:GA} displays the spin-resolved conductances as functions of $\lambda_{\Omega}$.
Generally, spin-up and spin-down conductances reach maxima when $\lambda_{\Omega}$ is small and decrease when $\lambda_{\Omega}$ is further enhanced.
Particularly, away from $\lambda_{\Omega}/\lambda_{SO} = 1.0$ the spin-down conductance of P configuration vanishes rapidly, while the spin-up conductance remains nonzero.
In the AP configuration, the decay of spin-up conductance is much faster than that of spin-down channel when $\lambda_{\Omega}/\lambda_{SO}$ is away from --1.0.
These leads to fully spin-polarized currents in certain ranges of $\lambda_{\Omega}$ as marked by gray regions on the figure.
In particular, P and AP configurations yield the positive and negative spin polarizations, respectively.

\begin{figure} [t]
  \includegraphics[width=9.0cm,height=6.0cm]{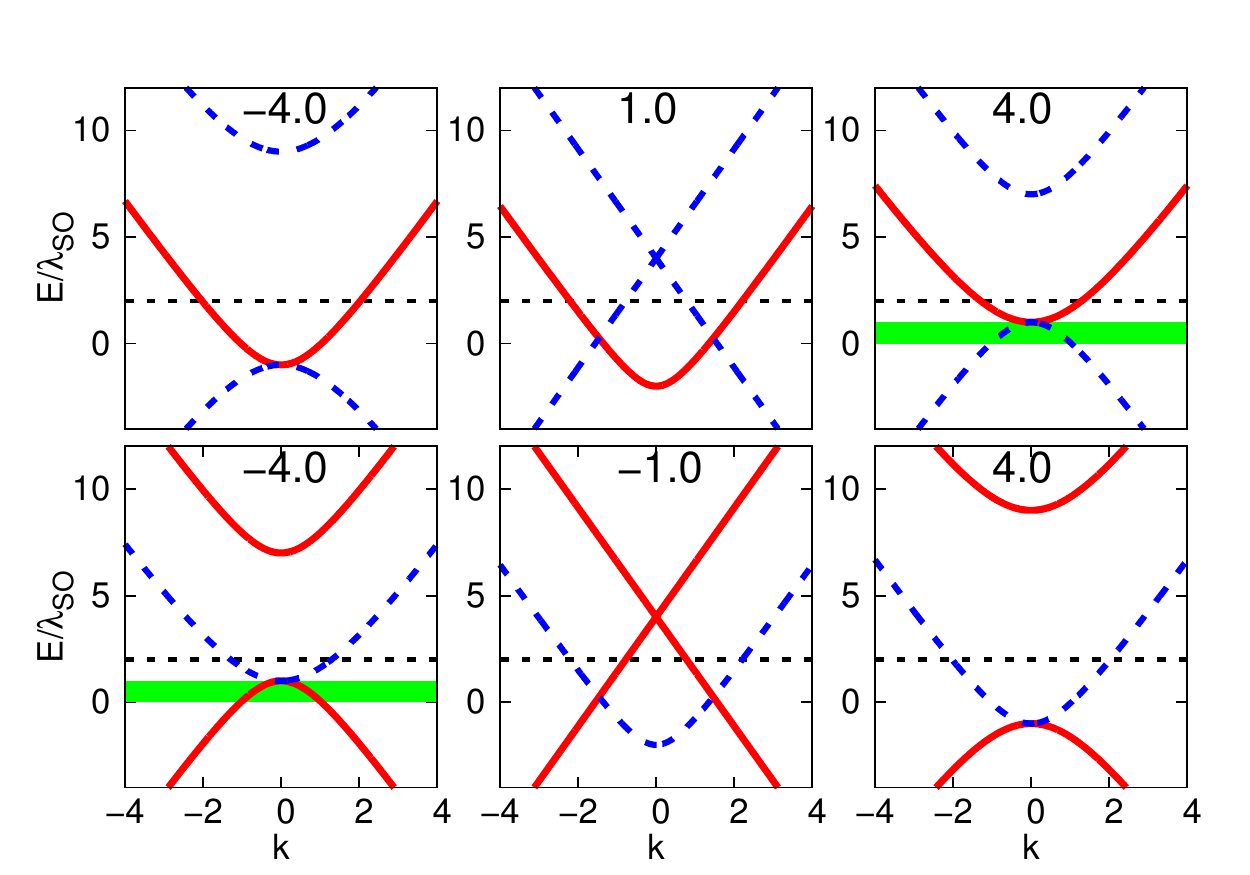}
  \caption{Low-energy band structures of third regions at $\lambda_\Omega/\lambda_{SO} = 1.0,\, \pm4.0$ for P configuration (upper row) and at $\lambda_\Omega/\lambda_{SO} = -1.0,\, \pm4.0$ for AP configuration (lower row). The solid red indicates the spin-up bands and the dash blue indicates spin-down bands.
    $h$ and $U$ are the same as in Fig~\ref{fig:GA}.
    The dot black line refers to $E/\lambda_{SO} = 2.0 $; and $E/\lambda_{SO} \in [0,1]$ is marked by the green region.}
  \label{fig:band1}
\end{figure}

The variations of spin-resolved conductances with $\lambda_{\Omega}$ can be clearly understood from low-energy band structures as shown in Fig. \ref{fig:band1}.
For P configuration (upper row), the spin-down band gap is closed at the value $\lambda_{\Omega}/\lambda_{SO} = 1.0$, therefore, conductance of this channel reaches maximum.
Away from this value, say $\lambda_{\Omega}/\lambda_{SO} = \pm 4.0$, the spin-down band gap is enlarged and the incident energy $E/\lambda_{SO} = 2.0$ (dotted black line) falls within the gap, while it crosses one spin-up band, leading to the full positive spin polarization as observed.
The explanation for AP configuration is similar to that of P configuration.
The spin-up band gap of AP configuration is, however, closed at $\lambda_{\Omega}/\lambda_{SO} = -1.0$.

\begin{figure} [h]
  \includegraphics[width=7.3cm,height=3.15cm]{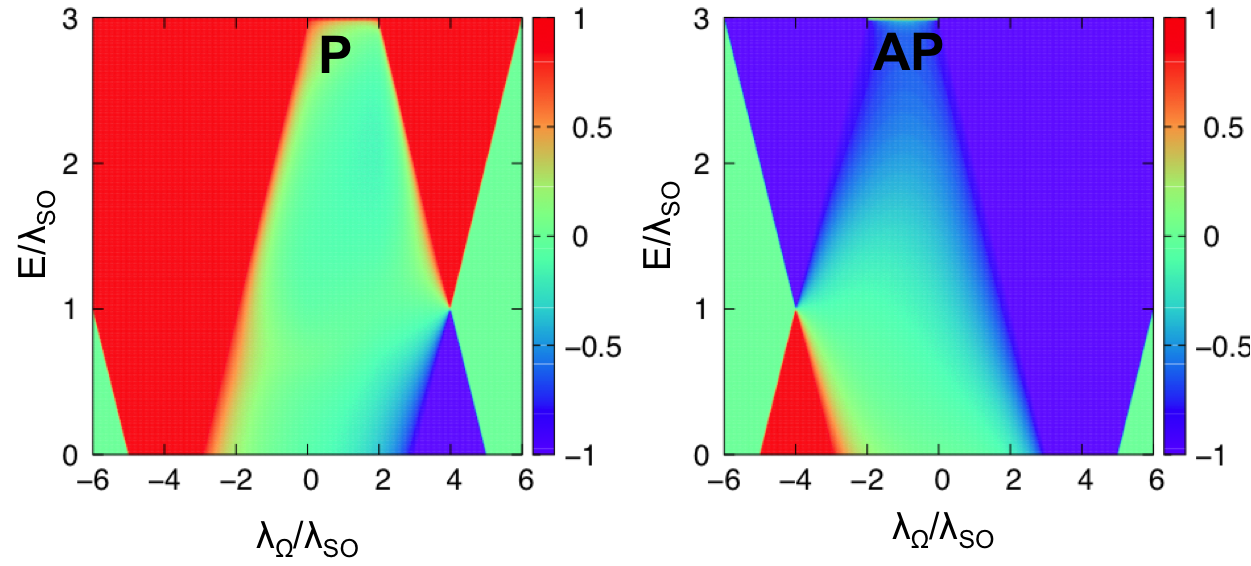}
  \includegraphics[width=7.5cm,height=3.25cm]{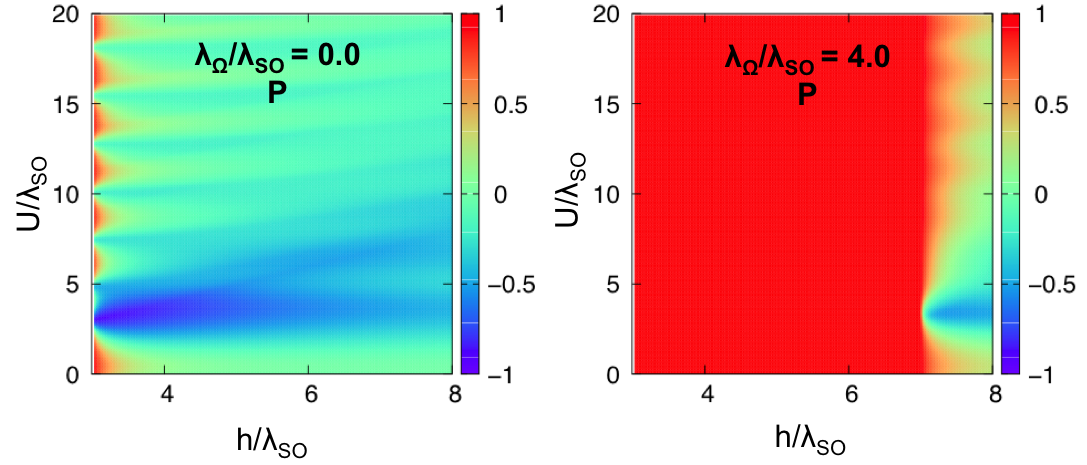}
  \caption{Upper panel: spin polarization $P_s$ as functions of $\lambda_{\Omega}$ and inicident energy $E$ for P (left) and AP (right) configurations with $h/\lambda_{SO} = 4.0$ and $U/\lambda_{SO} = 20.0$. Lower panel: spin polarization as functions of exchange field $h$ and barrier $U$ at $\lambda_\Omega/\lambda_{SO} = 0$ (left) and $4.0$ (right) for the P configuration with $E/\lambda_{SO} = 2.0 $.}
  \label{fig:ps_aehu}
\end{figure}

Obviously, the incident energy plays an important role to obtain the fully spin-polarized current.
In order to examine the interplay between the incident energy $E$ and $\lambda_{\Omega}$, we plot the spin polarizations of both P and AP configurations as functions of $E$ and $\lambda_{\Omega}$ on upper panel of Fig. \ref{fig:ps_aehu}.
For P configuration, the enhancement of incident energy under the right polarized light is resulted in the change of spin polarization from negative to positive.
When a left polarized light is applied, positive spin polarization irrespective of incident energy can be obtained.
For AP configuration, in contrast, the right polarized light yields negative spin polarization irrespective of incident energy, while the positive to negative transition occurs in the case of left polarized light.
The transition of spin polarization from negative to positive and {\it vice versa} can be demonstrated from low-energy band-structure as shown in Fig. \ref{fig:band1}.
We only consider the band structure of P configuration (upper row) at $\lambda_{\Omega}/\lambda_{SO} = 4.0$ and a similar explanation can be used for the AP configuration (lower row) at $\lambda_{\Omega}/\lambda_{SO} = -4.0$.
We can see that in the range $[0,1]$ marked by the green region, the incident energy will cross the spin-down band while it is within the gap of spin-up band, the negative spin polarization is then obtained.
On the other hand, when the incident energy increases out of the range, the spin-up density of state is nonzero while there is no density of states for spin-down channel, so that the spin polarization is positive.
In general, by changing circularly polarized light together with incident energy, a desired spin polarization can be easily realized in FNF silicene junction.

To further explore the enhancement of spin polarization under circularly polarized light, we plot, for example, the spin polarization of P configuration as functions of ferromagnetic exchange field $h$ and barrier height $U$ in lower panel of Fig. \ref{fig:ps_aehu}.
In the absence of light, the spin polarization is only occur when the exchange field $h$ is small.
Moreover, the spin polarization strongly oscillates with respect to barrier height $U$.
Interestingly, when the light is switched on, the full positive spin polarizations regardless of barrier height $U$ are observed in a large range of ferromagnetic exchange field $h$.

\begin{figure} [h]
  \includegraphics[width=8.5cm,height=3.25cm]{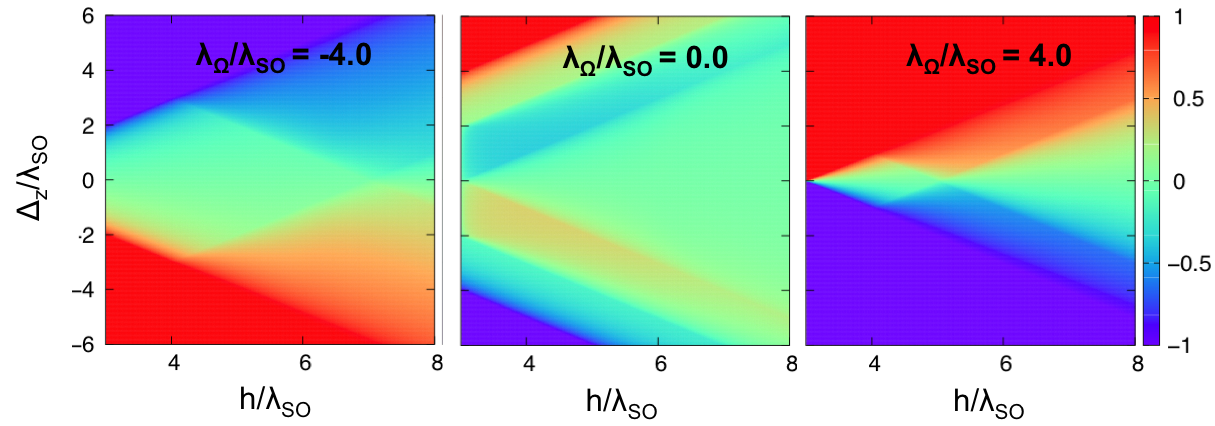}
  \caption{Valley polarization $P_v$ of P configuration as functions of $h$ and $\Delta_z$ at different values of $\lambda_\Omega$.
    The other parameters are: $E/\lambda_{SO} = 2.0 $ and $U/\lambda_{SO} = 20.0 $.}
  \label{fig:pv_eh}
\end{figure}

Let us now briefly discuss on the valley polarization.
The electric field applied on the third region is switched on to generate the valley dependent current.
Fig.~\ref{fig:pv_eh} shows the valley polarization $P_v$ as functions of staggered electric potential $\Delta_z$ and exchange field $h$.
Herein, we only consider the P configuration as an example, while the AP configuration can be done similarly.
Formally, fully valley-polarized current in FM silicene junction can be only realized when the staggered electric potential is larger than the exchange field, i.e. $\Delta_z/h > 1$ \cite{Saxena_prb2015}.
This argument is confirmed by the middle panel of the figure, where $\lambda_\Omega/\lambda_{SO} = 0$.
Once the light is switched on, the value of $\Delta_z$ to generate perfect valley polarization is substantially reduced, especially for the right polarized light.
Furthermore, the valley polarization is an odd function of $\lambda_\Omega$.
All these situations are easily understood from the second term of Hamiltonian (\ref{Heff}).
Since the circularly polarized light in the off-resonant regime is linearly dependent on the valley index $\eta$, the valley-degenerate breaking under staggered electric field will be significantly enhanced and the sign of valley polarization will vary accordingly to the direction of light.
Generally, the valley polarization in FNF silicene junction generated by staggered electric field will become more sensitive under a circularly polarized light.

\subsection{Tunneling magnetoresistance}

\begin{figure} [h]
\includegraphics[width=7.25cm,height=3.25cm]{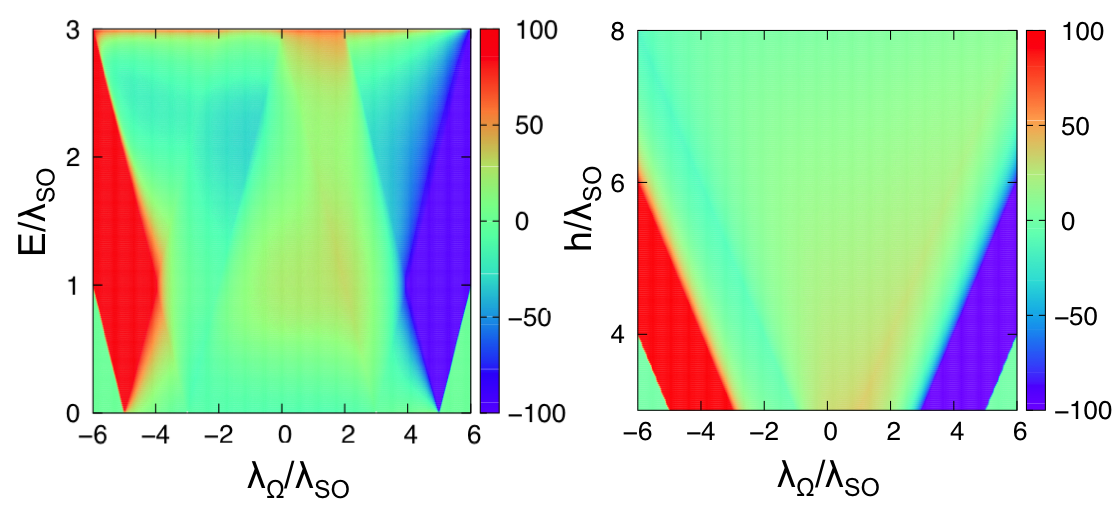}
\caption{2D contour plots of $TMR(\lambda_{\Omega},E)$ (left) and $TMR(\lambda_{\Omega},h)$ (right).
  The parameters used for left panel: $h/\lambda_{SO} = 4.0,\, U/\lambda_{SO} = 20.0 $; and for right panel: $E/\lambda_{SO} = 1.0,\, U/\lambda_{SO} = 20.0 $.
}
\label{fig:tmr_aeh}
\end{figure}

Finally, we consider the effect of circularly polarized light on TMR of the FNF silicene junction.
To eliminate the contribution from staggered potential to TMR, the external electric field is again turned off ($\Delta_z = 0$).
Left panel Fig. \ref{fig:tmr_aeh} shows TMR in \% as functions of incident energy $E$ and $\lambda_{\Omega}$.
100\% TMR can be achieved when $\left|\lambda_{\Omega}/\lambda_{SO}\right| > 4.0$.
Changing polarized direction of light from left to right gives rise to the transition from positive to negative TMR as previously observed in Ref.~\cite{Niu_epl2015}.
In contrast to Ref.~\cite{Wang_jap2013}, in which TMR of FNF silicene junction strongly oscillates with incident energy, we found that in the presence of circularly polarized light TMR is constant within large ranges of incident energy.

\begin{figure} [h]
\includegraphics[width=8.5cm,height=6.0cm]{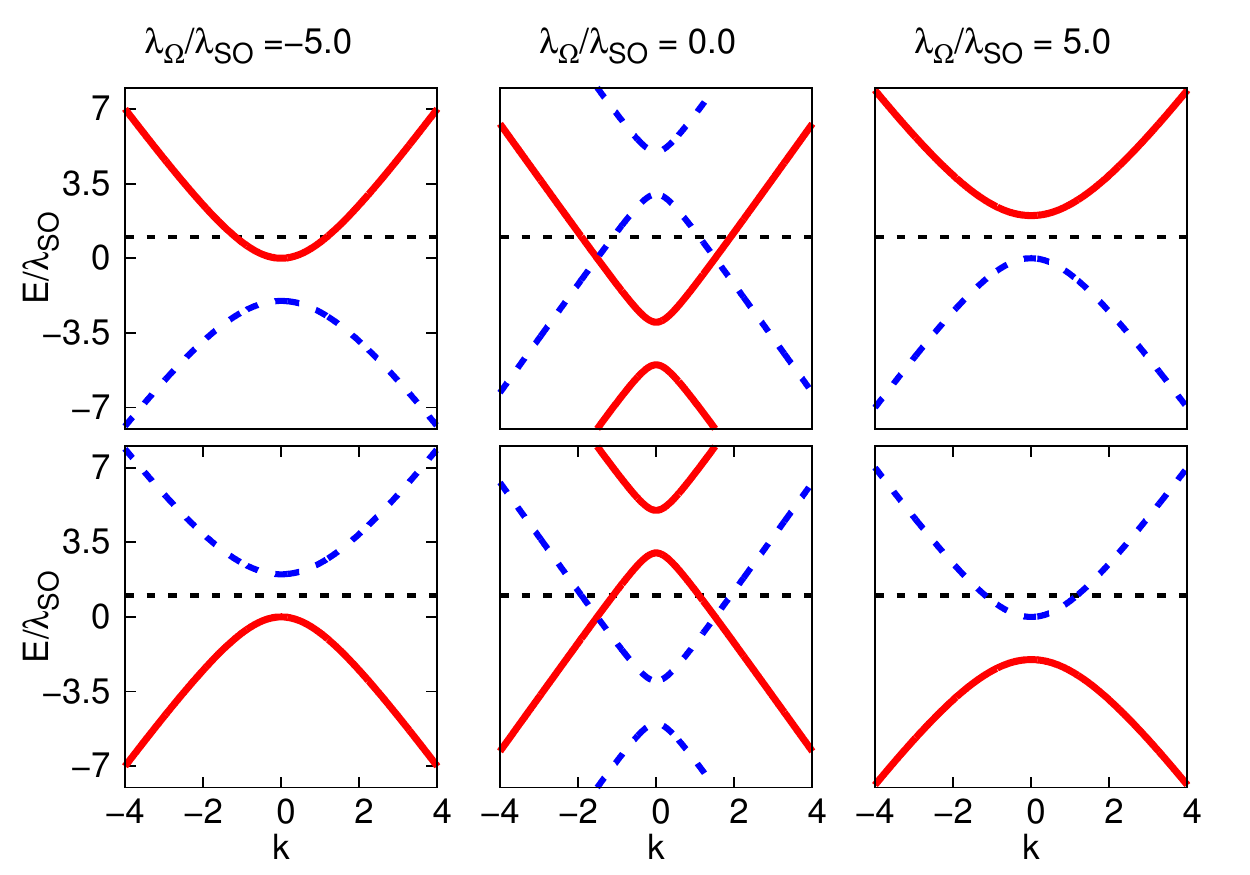}
\caption{Low-energy band structures at $\lambda_{\Omega}/\lambda_{SO} = -5.0$ (first column), $\lambda_{\Omega}/\lambda_{SO} = 0.0$ (second column), and $\lambda_{\Omega}/\lambda_{SO} = 5.0$ (last column) for P (upper row) and AP (lower row) configurations.
  The dot black line refers to $E/\lambda_{SO} = 1.0 $.
}
\label{fig:band2}
\end{figure}

Analyzing low-energy band structure of P and AP configurations as shown in Fig.~\ref{fig:band2} can provide the physical origin of giant TMR.
At $\lambda_{\Omega}/\lambda_{SO} = 5.0$, the incident energy is within the band gap for the P configuration, while it crosses  spin down band for AP configuration, leading to 100\% negative TMR.
At $\lambda_{\Omega}/\lambda_{SO} = -5.0$, the incident energy is within the band gap for the AP configuration and crosses the spin up band for P configuration, resulted in the 100\% positive TMR.
In general, there are two requirements to realize the 100\% TMR.
On one hand, the incident energy has to be located in the band gap for one configuration.
On the other hand, it has to cross one energy band for the other configuration.
These conditions can be easily derived from the eigenvalue of effective Hamiltonian (\ref{Heff}) in third region, $E = \pm \sqrt{\left(\hbar v_F k\right)^2+\left(\sigma \lambda_{SO}+\lambda_{\Omega}\right)^2} - \sigma h_3$ with $\left|h_3\right| = h$, as
\begin{align}
  \left(h - \lambda_{SO}\right) + E < \left|\lambda_\Omega\right| < \left(h + \lambda_{SO}\right) + E. \label{tmrcond}
  \end{align}
The condition (\ref{tmrcond}) is fulfilled by the right panel of Fig.~\ref{fig:tmr_aeh}, in which we plot TMR as functions of $\lambda_{\Omega}$ and $h$ at $E/\lambda_{SO} = 1$.

\begin{figure} [t]
\includegraphics[width=8.5cm,height=3.15cm]{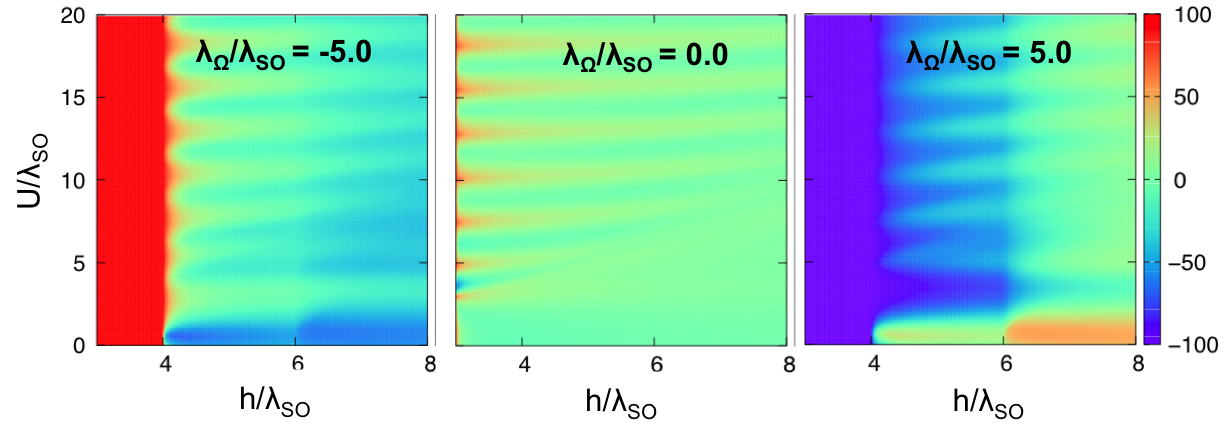}
\caption{ TMR (in \%) as functions of exchange field $h$ and barrier $U$ at different values of $\lambda_{\Omega}$. The incident energy is $E/\lambda_{SO} = 1.0 $.}
\label{fig9}
\end{figure}

It is common that TMR will strongly oscilate with respect to $U$  \cite{Wang_jap2013,Saxena_prb2015} and a significant TMR value can be only obtained when the exchange field $h$ is smaller than the barrier $U$ \cite{Saxena_prb2015}.
Consequently, it is not easy to get a desired TMR value in practice.
It is therefore interesting and informative to examine the variation of TMR with the exchange field $h$ and barrier $U$ in presence of circularly polarized light.
The results are summarized in Fig.~\ref{fig9}.
As expected, in the absence of circularly polarized light, TMR strongly oscilates with respect to $U$ and only has the significant value at small $h$.
Nevertheless, when the junction is exposed to a left (right) circularly polarized light, 100\% positive (negative) TMR can be achieved regardless of $U$.
More importantly, the junction can provide 100\% TMR in a large range of $h$ even when $U$ vanishes.

\section{Conclusion}
In summary, the influence of a circularly polarized light in off-resonant regime on ballistic transport through a FNF silicene junction has been systematically investigated.
We have shown that the spin/valley polarizations and TMR are significantly enhanced under a circularly polarized light.
Particularly, tuning incident energy in the presence of a circularly polarized light leads to the transition of spin polarization from positive to negative and {\it vice versa}.
Thanks to the valley dependence of off-resonant circularly polarized light, the perfect valley polarization can be achieved even when staggered electric field is much smaller than exchange field.
The most importantly, the perfect spin polarization and 100\% TMR irrespective of barrier $U$ can be realized when the junction is exposed to a circularly polarized light.
We have also gained a condition for observing the 100\% TMR.

\section*{Acknowledgment}
This work is supported by Vietnamese National Foundation of Science and Technology Development (NAFOSTED) under Grant No. 103.01-2015.14. 

\bibliography{main}

\end{document}